\documentclass[12pt]{article}
\usepackage{graphics}
\usepackage{amssymb,epsfig,amsmath,euscript,array}
\usepackage{axodraw4j}
\usepackage{subfigure}
\usepackage{color}
\usepackage{cite}

\makeatletter
\@addtoreset{equation}{section}
\makeatother

\definecolor{holger}{rgb}{0,0.5,0.7}
\definecolor{edit}{rgb}{1,0,0}

\definecolor{durbeer}{rgb}{1,0,0}

\definecolor{durbeer2}{rgb}{0.8,0,0.5}

\newcommand{\Tr}{\text{Tr}}

\newcommand{\Aslash}{A\!\!\!/\,_{1}}
\newcommand{\Bslash}{A\!\!\!/\, _{2}}

\newcounter{multieqs}
\newcommand{\ii}{{\rm i}}

\def\bd{\begin{document}}
\def\ed{\end{document}}
\def\nn{\nonumber}
\def\bea{\begin{eqnarray}}
\def\eea{\end{eqnarray}}
\let\bm=\bibitem
\let\la=\label

\begin{document}

\hfill{IPPP/09/36; DCPT/09/72}\\[-0.9cm]

\vspace{20pt}

\begin{center}

{\LARGE \bf Magnetic Mixing}\\
{\bf --}\\
{\Large\bf Electric Minicharges from Magnetic Monopoles} \\[1.5ex]

\vspace{30pt}

{\bf  Felix Br\"ummer, Joerg Jaeckel and Valentin V. Khoze}

{\small \em {Institute for Particle Physics Phenomenology, \\
Department of Physics, Durham
University,\\ Durham DH1 3LE, United Kingdom}}

\vspace{10pt}

{\sffamily \tt
felix.bruemmer@durham.ac.uk\\
joerg.jaeckel@durham.ac.uk
valya.khoze@durham.ac.uk}

\vspace{30pt}
\end{center}

\begin{abstract}
Many extensions of the Standard Model require the existence of a
``hidden'' sector. We consider settings where the hidden sector in the infrared contains a U(1) gauge factor with magnetic monopoles, for
instance 't Hooft-Polyakov monopoles of an underlying non-abelian gauge group.
In the presence of CP violation these monopoles acquire an electric charge in the hidden sector due to the Witten effect.
We show that quite generally they also acquire (small) electric charges under the visible electromagnetic gauge group.
This is a result of ``magnetic mixing'' which, as we show, often arises as a natural partner of kinetic mixing.
Both kinetic and magnetic mixing are naturally induced radiatively even if the low-energy U(1)s arise from a single non-abelian gauge group.
We argue that the hidden sector monopoles can be light and their electric minicharges could thus be testable
in current and near future low-energy experiments.
\end{abstract}

\setcounter{page}{0}
\thispagestyle{empty}
\newpage


\section{Introduction}

It is well known that in gauge theories whose low-energy description contains multiple U(1) gauge factors, kinetic terms
which mix field-strengths of different U(1)s
are allowed by both Lorentz symmetry and gauge invariance. Such kinetic mixing terms can lead to the appearance of
particles with tiny electric charges \cite{Holdom:1985ag} if one of the U(1)s is identified with the electromagnetic U(1) of the Standard Model
and the others belong to a hidden sector. It is much less familiar that also non-diagonal $\theta$-terms are possible.
For example, in a model with gauge group U$(1)_1\times$ U$(1)_2$ the most
general renormalizable Lagrangian in the gauge sector is\footnote{A note on conventions: for uniformity of notation we use the $\frac{1}{e^2}$ normalisation in front of $F^{\mu\nu} F_{\mu\nu}$
terms even in the abelian theory. The couplings of the abelian as well as of the underlying non-abelian theories are
denoted as $e$. The symbol $g$ is reserved for magnetic charges of monopoles.}
\begin{equation}
\label{lagrangian}
\begin{split}
{\cal L}=-\frac{1}{4}\Bigl\{\,&\frac{1}{e_1^2}\,F^{\mu\nu}_{1} F_{1,\mu\nu}\,
+\,\frac{1}{e_2^2}\,F^{\mu\nu}_{2} F_{2,\mu\nu}\,+\,\frac{2\chi}{e_1e_2}\, F^{\mu\nu}_{1}F_{2,\mu\nu}\\
&+\,\frac{1}{8\pi^2}\left(\theta_{11}\,F^{\mu\nu}_{1}\tilde F_{1,\mu\nu}\,+\,\theta_{22}\,F^{\mu\nu}_{2}\tilde F_{2,\mu\nu}\,
+\,2\theta_{12}\,F^{\mu\nu}_{1}\tilde F_{2,\mu\nu}\right)\Bigr\}.
\end{split}
\end{equation}
Here, the $F^{\mu\nu}_{I}$ are the field strengths of U$(1)_{I}$ (with couplings constants $e_{I}$).
Their duals are defined as usual:
\begin{equation}
\tilde F^{\mu\nu}_{I}=\frac{1}{2}\epsilon^{\mu\nu\kappa\lambda}F_{I,\kappa\lambda}\,.
\end{equation}

The standard kinetic terms as well as the
$\theta_{11}$ and $\theta_{22}$ $F\tilde F$-terms are the diagonal entries
in the Lagrangian \eqref{lagrangian}. Mixing between the different U(1) sectors is represented by
the off-diagonal
$\chi$-term and $\theta_{12}$-term.

The $\chi$-term in \eqref{lagrangian} is the familiar kinetic mixing term. The term $\propto \theta_{12} {\bf E_1}\cdot{\bf B_2}$
is a total derivative. It represents
the mixing of the magnetic field of the second U(1) with the electric field of the first.
We will refer to it as ``magnetic mixing''.

While their presence is important in the non-abelian case,
$\theta$-terms are usually ignored in abelian theories since they do not affect the equations of motion and carry no topological charges.
A well-known exception is provided by the Witten effect~\cite{Witten:1979ey}, where
the $\theta F\tilde{F}$-term induces electric charges for magnetic monopoles.
This effect is difficult to observe for monopoles of ordinary electromagnetism,
since their masses are extremely large\footnote{For a Grand Unified Theory the monopole
mass would be $\sim M_{\rm GUT}/e_{\rm GUT}^2\gg M_{\rm GUT}$.}.
But when two or more  U$(1)$ factors become involved, we will argue
in Sect.~\ref{witten} that the effects of $\theta$-terms do become phenomenologically
interesting if there is a
light monopole associated with the hidden sector\footnote{Magnetic monopoles and electric minicharges acquired via kinetic mixing
are mutually compatible, because a Dirac-type quantization condition continues to hold \cite{Brummer:2009cs}.} .
This monopole can acquire a fractional electric charge under the Standard Model U$(1)$
either due to magnetic mixing or due to
an ordinary $\theta$-term in the hidden sector along with nonzero kinetic mixing.
Through their Standard Model charges, light hidden monopoles might then be detectable
in the ongoing and future experimental searches for minicharged particles
\cite{Jaeckel:2009dh,Davidson:2000hf,Gies:2006hv,Badertscher:2006fm,Gninenko:2006fi,Ahlers:2007rd,Cameron:1993mr,Gies:2009wx,Gies:2006ca,Zavattini:2007ee,Dobrich:2009kd,Melchiorri:2007sq,Ahlers:2009kh,Gluck:2007ia}.

It is important to note that for our purposes the hidden sector gauge group does not necessarily have to contain a  U$(1)$ at the
fundamental level. In fact, our prime example will be based on a non-abelian hidden sector gauge group,
broken to U$(1)$s by an adjoint Higgs field. While magnetic monopoles may be added or left out at will in a pure U$(1)$, the non-abelian model
necessarily contains 't Hooft-Polyakov monopoles \cite{'tHooft:1974qc,Polyakov:1974ek}. As opposed to their pure U$(1)$ (Dirac) counterparts
they are non-singular and their properties are calculable.
Kinetic mixing between an unbroken U$(1)$ subgroup in the hidden sector and the Standard Model
can be induced radiatively\footnote{See \cite{Dienes:1996zr,Batell:2005wa,Abel:2006qt} for some examples in field and string theory model building.}; as we will
show in Sect.~\ref{radiative}, the same is also true for magnetic mixing.

For concreteness, let us assume that a hidden sector monopole emerges from a spontaneously broken non-abelian gauge
group. There are then two distinct scenarios in which it can be light. In the first case, one uses the semi-classical estimate for the monopole mass, $m\approx v/e$ (where $v$ is the breaking scale, and $e$ is the gauge coupling). By sufficiently lowering $v$, the monopole can be light even at weak coupling\footnote{The semi-classical
approximation is valid for $\Lambda<ev<v/e$,
where $\Lambda$ is the dynamical transmutation scale of the
hidden sector non-abelian gauge group.}.

A more interesting scenario for having light monopoles arises in the context of a strongly coupled hidden sector.
A naive semi-classical approximation suggests that the monopole mass $\sim v/e$ can be small.
The fact that monopoles do indeed become light in the strongly coupled effective theory in the infrared was famously demonstrated by Seiberg and Witten
in the calculable context of ${\cal N}=2$ supersymmetric QCD~\cite{Seiberg:1994rs, Seiberg:1994aj}.
We expect that qualitative features of the Seiberg-Witten model, in particular the fact that the monopoles become light when the original non-abelian theory becomes strongly coupled,
hold in more general
situations. Imagine that the hidden sector is described by an SU($N$) gauge theory with an adjoint Higgs which breaks the gauge group to U(1)$^{N-1}$.
This theory contains (a) magnetic monopoles and (b) has a low-energy description in terms of U(1) degrees of freedom. In the weakly coupled regime
$v\gg\Lambda$ the monopoles
are heavy. In analogy to the Seiberg-Witten case we expect that even at strong coupling, i.e.~where the vev is not much greater than $\Lambda$, the low-energy
dynamics is still described by the U(1)$^{N-1}$ theory.
This is a strongly coupled theory which has a weakly coupled magnetic dual description with the original
magnetic monopoles appearing as light elementary fields.

The paper is set up as follows. In the following Sect.~\ref{witten} we will briefly review the Witten effect and demonstrate that, when we generalize it
to two U(1) factors, the monopoles of one U(1) can acquire electric charges under the other U(1). We will then use this result in Sect.~\ref{interaction}
to provide a general formula for the interaction energy between two static sources with electric and monopole charges which is valid
in the presence of both kinetic mixing and $\theta$-terms.
This can be used to define ``charge'' by a measurable quantity: the energy. Then in Sect.~\ref{radiative} we will discuss how
kinetic and magnetic mixing (as well as diagonal $\theta$-terms) in the low-energy abelian theory
arise radiatively.
We will show this in the cases where (a) the hidden sector gauge group was distinct from the visible sector, and
(b) where all U(1) factors originated from a single non-abelian gauge theory. We will summarize and conclude in Sect.~\ref{conclusions}.

\section{The Witten effect and  magnetic mixing}\label{witten}
Witten has shown \cite{Witten:1979ey} that a $\theta$-term gives fractional electric charges to magnetic monopoles.
First we briefly recall a simple version of the derivation which is due to Coleman \cite{Coleman:1982cx}.
In a model with gauge group U$(1)$, the Lagrangian density is
\begin{equation}
{\cal L}=-\frac{1}{4 e^2} F^{\mu\nu}F_{\mu\nu}-\frac{\theta}{32\pi^2} F^{\mu\nu}\tilde F_{\mu\nu}.
\end{equation}
The $\theta$-term can be written in terms of the magnetic and electric fields,
\begin{equation}
-\frac{\theta}{32\pi^2} F^{\mu\nu}\tilde F_{\mu\nu}=\frac{\theta}{8\pi^2}\,{\bf E}\cdot{\bf B}.
\end{equation}
In a magnetic monopole background with a superimposed static electromagnetic potential \mbox{$(A^\mu)=(A^0,{\bf A})$}, the field strengths are
\begin{equation}
{\bf E}=\nabla A^0
,\qquad {\bf B}=\nabla\times{\bf A}+\frac{ge}{4\pi}\frac{{\bf r}}{r^3},
\end{equation}
where $g$ is the monopole charge. The scalar potential $A^0$ then appears in the Lagrangian for the $\theta$-term as follows:
\begin{eqnarray}
L_\theta = \frac{\theta }{8\pi^2}\int d^3r\,{\bf E}\cdot{\bf B}\!\!&=&\!\!
\frac{\theta }{8\pi^2}\int d^3r\,(\nabla A^{0})\cdot(\nabla\times{\bf A}+\frac{ge}{4\pi}\frac{{\bf r}}{r^3})
\\\nonumber
\!\!&=&\!\! -\frac{\theta e g}{32\pi^3}\int d^3r\,A^0\,\nabla\cdot\frac{{\bf r}}{r^3}=-\frac{\theta e g}{8\pi^2}\int d^3r\, A^0\,\delta^3({\bf r})
\end{eqnarray}
The right hand side corresponds to the interaction of an electric point charge $-\theta e g/(8\pi^2)$ at the origin, i.e.~at the location of the monopole,
with the electrostatic potential $A^{0}$. Using the Dirac quantization condition\footnote{In our conventions half-integer
electric charges are allowed.} $eg=4\pi$ the magnetic monopole has acquired
an electric charge $-\theta /(2\pi)$.
A more precise argument, which does not involve singular charge densities, can be given for monopoles arising from spontaneously
broken non-abelian gauge groups \cite{Witten:1979ey}.

It is now straightforward to generalize this to the case of multiple U(1) gauge factors and off-diagonal $\theta$-terms.
For example consider the case of two U(1)s with a Lagragian density given by Eq.~\eqref{lagrangian}.
The diagonal $\theta_{11}$ and $\theta_{22}$ will lead to electric charges of the monopole-1 under U(1)$_{1}$ and the monopole-2 under U(1)$_{2}$, respectively.
However, we can also have a ``mixing'' $\theta_{12}$-term. Let us check how this affects a monopole-2. In a static situation with a monopole-2 the
electric and magnetic fields of U(1)$_{1}$ and U(1)$_{2}$ read
\begin{equation}
{\bf E_{1}}=\nabla A^{0}_{1}
,\qquad {\bf B_{1}}=\nabla\times{\bf A}_{1},\qquad{\bf E_{2}}=\nabla A^{0}_{2}
,\qquad {\bf B_{2}}=\nabla\times{\bf A}_{2}+\frac{e_{2} g_{2}}{4\pi}\frac{{\bf r}}{r^3}.
\end{equation}
We then have
\begin{eqnarray}
L_{\theta_{12}}\!\!&=&\!\!\frac{\theta_{12}}{8\pi^2}\int d^{3}x\,({\bf E}_{1}\cdot{\bf B}_{2}+{\bf E}_{2}\cdot{\bf B}_{1})
=-\frac{\theta_{12}}{8\pi^2}\int d^{3}x\, A^{0}_{1}\nabla\cdot\left(\frac{g_{2}e_{2}}{4\pi} \frac{{\bf r}}{r^3}\right)
\\\nonumber
\!\!&&\!\!\qquad\qquad\qquad\qquad\qquad\qquad=-\frac{\theta_{12}e_{2}g_{2}}{8\pi^2}\int d^{3} x\, A^{0}_{1}\delta^{3}({\bf r})
=-\frac{\theta_{12}}{2\pi}\int d^{3} x\, A^{0}_{1}\delta^{3}({\bf r}),
\end{eqnarray}
where we have used the quantization condition $e_{2}g_{2}=4\pi$ in the last step. At this point we stress that this quantization condition also holds
in the case of multiple U(1)s, even with kinetic mixing (cf.~\cite{Brummer:2009cs} for details).
This shows that the magnetic monopole-2 has acquired an electric charge $-\theta_{12}/(2\pi)$ under the group U(1)$_{1}$.

In the specific case that U(1)$_{1}$ is the electromagnetic gauge group of the Standard Model and U(1)$_{2}$ is a hidden sector gauge group
this means that the formerly invisible hidden monopole has acquired an electric charge and couples to the electromagnetic photon.

In the case of multiple U(1)s with magnetic mixing, the electric and magnetic
charges $Q^{e}$ and $Q^{m}$ are given by
\begin{eqnarray}
\label{charges}
Q^{e}_{I}\!\!&=&\!\! n^{e}_{I}-\frac{\theta_{IJ}}{2\pi}n^{m}_{J},
\\\nonumber
Q^{m}_{I}\!\!&=&\!\!4\pi n^{m}_{I},
\end{eqnarray}
where $n^{e}$ and $n^{m}$ are the electric and magnetic (monopole) numbers, respectively.

\section{Interaction energy between charged particles}\label{interaction}
In the previous section we have determined the charges for particles with both electric and magnetic quantum numbers.
However, in presence of kinetic mixing the physical meaning of charge is not completely straightforward since particles electrically charged
under one U(1) can interact with the particles of another U(1) via the kinetic mixing term.
As we will briefly review below the kinetic term can be diagonalized by a linear transformation of the gauge fields. This, however, redefines the charges.
It is more appropriate to consider the interaction energy between two charged particles, which is an observable quantity and therefore basis-independent.
Below we will provide a general formula for this energy.
One can then measure the ``charges'' by measuring the interaction energy with respect to reference particles.

Let us briefly recall how effective minicharges arise for electrically charged particles in the context of kinetic mixing.
The kinetic terms in Eq.~\eqref{lagrangian} can be diagonalized by a shift
\begin{equation}
\label{shift}
A^{\mu}_{2}\rightarrow A^{\mu}_{2}-\frac{\chi e_{2}}{e_{1}} A^{\mu}_{1}.
\end{equation}
Apart from a multiplicative renormalization of the gauge coupling,
\begin{equation}
\label{rescaling}
e^{2}_{1}\rightarrow e^{2}_{1}/(1-\chi^2),
\end{equation}
the $1$-fields remain
unaffected by this shift. Consider now for example a hidden fermion $f$ charged under $A^{\mu}_{2}$.
Applying the shift~\eqref{shift} to the coupling term, we find:
\begin{equation}
\bar{f}\Bslash\, f\rightarrow \bar{f}\Bslash\, f-\frac{\chi e_{\rm 2}}{e_{1}}\bar{f}\Aslash\,
 f.
\end{equation}
Since the kinetic term is now diagonal, it is clear that the particle $f$ (which was originally charged
only under U(1)$_{2}$) interacts with the U(1)$_{1}$ gauge field and appears to have a charge $-\chi e_{2}/e_{1}$.

In principle one could now diagonalize the kinetic terms in the general case and then define the ``charges'' with respect to the gauge fields in this basis.
A somewhat more physical (and basis independent) approach is to concentrate directly on the measurable interaction energy between two particles.

It will be convenient for our purposes to define a $\tau$-matrix of complexified coupling constants of the U(1)$^{k}$ low-energy theory,
\begin{equation}
\tau_{IJ}=\ii(\tau_{2})_{IJ}+(\tau_{1})_{IJ}=\left(\frac{4\pi\ii}{e^{2}}\right)_{IJ}+\left(\frac{\theta}{2\pi}\right)_{IJ},
\end{equation}
where $(1/e^{2})_{IJ}$ is the matrix of the inverse gauge couplings (the off-diagonal terms give kinetic mixing).
For example, for the Lagrangian Eq.~\eqref{lagrangian} we have
\begin{equation}
\label{deftau}
\tau_{IJ}= 4\pi\ii \left(
                \begin{array}{cc}
                  \frac{1}{e^{2}_{1}} & \frac{\chi}{e_{1}e_{2}} \\
                  \frac{\chi}{e_{1}e_{2}} & \frac{1}{e^{2}_{2}} \\
                \end{array}
              \right)+\frac{1}{2\pi}\left(
                                      \begin{array}{cc}
                                        \theta_{11} & \theta_{12} \\
                                        \theta_{12} & \theta_{22} \\
                                      \end{array}
                                    \right).
\end{equation}

To obtain the interaction energy let us start by deriving the static
electric potential for an electrically charged particle.
From the Lagrangian Eq.~\eqref{lagrangian} and using our definition
\eqref{deftau} the equation of motion for $\phi_{I}=A^{0}_{I}$ is,
\begin{equation}
\label{eom}
({\rm Im}\,\tau)_{IJ}\Delta \phi_{J}=Q^{e}_{I}\,4\pi\,\delta^{3}({\bf r}).
\end{equation}
The appearance of the imaginary part is due to the fact that the $\theta$-terms, i.e.~the real part of $\tau$, multiply a total derivative in the Lagrangian
and therefore do not affect the equations of motion.
The equation of motion Eq.~\eqref{eom} can be solved, yielding
\begin{equation}
\phi_{I}=\frac{1}{r}\,\left(\frac{1}{{\rm Im} \tau}\right)_{IJ}Q^{e}_{J}.
\end{equation}
The interaction energy of an electric probe charge $Q^{e\,\prime}$ in this potential is then
\begin{equation}
E^{\rm electric}_{\rm Int}=\frac{1}{r}\,Q^{e\,\prime}_{\rm I}\phi_{I}=\frac{1}{r} Q^{e\,\prime}_{I}\left(\frac{1}{{\rm Im \tau}}\right)_{IJ}Q^{e}_{J}.
\end{equation}
Using Eq.~\eqref{charges} to relate the electric charge to the electric and magnetic monopole numbers we find
\begin{equation}
\label{int1}
E^{\rm electric}_{\rm Int}=\frac{1}{r}\,(n^{\prime}_{e}-\tau_{1}n^{\prime}_{m})_{I}\left(\frac{1}{\tau_{2}}\right)_{IJ}(n_{e}-\tau_{1}n_{m})_{J}.
\end{equation}
A similar procedure can be performed for the magnetic charges and the magnetic interaction energy. The only subtlety is in obtaining the
correct normalization for the magnetic charges. In our conventions this requires a factor of ${\rm Im}\tau$ for the magnetic charges.
Accordingly we obtain
\begin{equation}
\label{int2}
E^{\rm magnetic}_{\rm Int}=\frac{1}{r}\,n^{\prime}_{m,I}(\tau_{2})_{IJ}\left(\frac{1}{\tau_{2}}\right)_{JK}(\tau_{2})_{KL}n_{m,K}.
\end{equation}

Eqs.~\eqref{int1} and \eqref{int2} can then be combined to give the total interaction energy between two particles with electric and magnetic
numbers, $n^{e},\, n^{e\,\prime}$ and $n^{m},\, n^{m\,\prime}$,
\begin{eqnarray}
\label{total}
E^{\rm total}_{\rm Int}\!\!&=&\!\!\frac{1}{r}\,
(n^{e\,\prime\,T},n^{m\,\prime\,T})
\left(
\begin{array}{cc}
1 & 0 \\
-\tau_{1} & \tau_{2} \\
\end{array}
\right)
\left(
\begin{array}{cc}
\frac{1}{\tau_{2}} & 0 \\
1 & \frac{1}{\tau_{2}} \\
\end{array}
\right)
\left(
\begin{array}{cc}
1 & -\tau_{1} \\
0 & \tau_{2} \\
\end{array}
\right)
\left(
  \begin{array}{c}
    n^{e} \\
    n^{m} \\
  \end{array}
\right)
\\\nonumber
\!\!&=&\!\!\frac{1}{r}\,
(n^{e\,\prime\,T},n^{m\,\prime\,T})
\left(
\begin{array}{cc}
\tau^{-1}_{2} & -\tau^{-1}_{2}\tau_{1} \\
-\tau_{1}\tau^{-1}_{2} & \tau_{1}\tau^{-1}_{2}\tau_{1}+\tau_{2} \\
\end{array}
\right)
\left(
  \begin{array}{c}
    n^{e} \\
    n^{m} \\
  \end{array}
\right)
.
\end{eqnarray}
This expression has many nice features. In particular, it is
invariant under basis changes for the gauge fields. Moreover, it is
also invariant under Sp$(2k,{\mathbf Z})$ duality transformations,
generated by exchanging the electric
and the magnetic degrees of freedom and by shifting the
$\theta$-terms by $2\pi$.
These duality transformations act as
\begin{equation}
\label{dual}
\tau\,\rightarrow\,(A\tau+B)(C\tau+D)^{-1},\quad
\left(
\begin{array}{c}
n^{e} \\
n^{m} \\
\end{array}
\right)\rightarrow
M \left(
\begin{array}{c}
n^{e} \\
n^{m} \\
\end{array}
\right)
,\quad M=\left(
            \begin{array}{cc}
              A & B \\
              C & D \\
            \end{array}
\right)\in {\rm Sp}(2k,{\mathbf Z}),
\end{equation}
where $A,B,C,D$ are $k\times k$ integer matrices with
\begin{equation}
\begin{split}
AB^{T}=BA^{T},\quad B^{T}D=D^{T}B,&\quad A^{T}C=C^{T}A,\quad D^{T}C=C^{T}D,\\
 A^{T}D-C^{T}B=&AD^{T}-BC^{T}=\mathbf{1}.
\end{split}
\end{equation}
For a more detailed account of Sp$(2k,{\mathbf Z})$ duality, see the Appendix.

To clarify the physical meaning let us look at some specific examples with two U(1) gauge groups.
The first U(1) we will assume to be the ordinary electromagnetic U(1) with an electron. The second U(1) we will take to be a ``hidden'' U(1)
with electrically and magnetically charged particles.

\noindent {\bf Example 1:} Our first example is actually the standard kinetic mixing already discussed at the beginning of this section.
If all $\theta$-terms vanish, $\tau_{1}=0$, the only possible source of interaction between the two U(1)s is the kinetic mixing $\chi$.
Let us look at the interaction between an electron, $n^{e}=(-1,0)^{T},\,n^{m}=(0,0)^{T}$ and a hidden electrically charged particle $n^{e}=(0,1)^{T},\,n^{m}=(0,0)^{T}$.
Plugging this into Eq.~\eqref{total} and denoting $(\tau_{2})_{12}=4\pi\chi/(e_{1}e_{2})$ as in Eq.~\eqref{deftau}, we find
\begin{equation}
E_{\rm Int}=\frac{1}{4\pi r}\frac{\chi e_{1} e_{2}}{1-\chi^2}.
\end{equation}
This is exactly the interaction energy we expect for a particle of visible electric charge $-\chi e_{2}/e_{1}$ (as always in units of $e_{1}$) as anticipated from the
discussion of kinetic mixing. The expression indeed also automatically accounts for the renormalization of the electric charge by a factor of $1/\sqrt{1-\chi^2}$.
The latter can be read off from the fact that the interaction energy between two electrons is
\begin{equation}
E_{\rm Int}=\frac{1}{4\pi r}\frac{e^{2}_{1}}{1-\chi^2}.
\end{equation}
A brief calculation shows that the interaction energy between an electron and a hidden monopole $n^{e}=(0,0)^{T},\,n^{m}=(0,1)^{T}$ indeed vanishes
as expected from the discussion in \cite{Brummer:2009cs}.

\noindent {\bf Example 2:} We are now ready to look at a situation with a non-trivial magnetic mixing.
We take $\theta_{12}\neq 0$ and set $\theta_{11}=\theta_{22}=\chi=0$.
The interesting interaction is now the one between an electron and a hidden monopole, resulting in
\begin{equation}
E_{\rm Int}=\frac{1}{4\pi r}\frac{\theta_{12}e^{2}_{1}}{2\pi},
\end{equation}
exactly as one would expect for a particle with ordinary electric charge $-\theta_{12}/(2\pi)$.

In this simple situation it seems that the hidden monopole behaves very similar to the ordinary hidden electric particle in our first example with kinetic mixing.
For the simple situation discussed here this is indeed the case as one can see by explicitly performing an electric magnetic duality transformation for the hidden
sector U(1) factor according to Eq.~\eqref{dual},
\begin{equation}
A=D=\left(
    \begin{array}{cc}
      1 & 0 \\
      0 & 0 \\
    \end{array}
\right),\quad
B=-C=\left(
    \begin{array}{cc}
      0 & 0 \\
      0 & 1 \\
    \end{array}
\right).
\end{equation}
This duality transformation changes,
\begin{equation}
\tau=
\left(
  \begin{array}{cc}
    \frac{4\pi \ii}{e^{2}_{1}} & \frac{\theta_{12}}{2\pi} \\
    \frac{\theta_{12}}{2\pi} & \frac{4\pi \ii}{e^{2}_{2}} \\
  \end{array}
\right)\stackrel{M}{\longrightarrow}
\left(
  \begin{array}{cc}
    \frac{4\pi \ii}{e^{2}_{1}} +\frac{\ii e^{2}_{2}\theta^{2}_{12}}{16\pi^3}& \frac{\ii e^{2}_{2}\theta_{12}}{8\pi^2} \\
    \frac{\ii e^{2}_{2}\theta_{12}}{8\pi^2} & \frac{\ii e^{2}_{2}}{4\pi} \\
  \end{array}
\right),\qquad \left(
                 \begin{array}{c}
                   (n^{e}_{1},n^{e}_{2})^{T} \\
                   (n^{m}_{1},n^{m}_{2})^{T} \\
                 \end{array}
               \right)\stackrel{M}{\longrightarrow}
\left(\begin{array}{c}
                   (n^{e}_{1},n^{m}_{2})^{T} \\
                   (n^{m}_{1},n^{e}_{2})^{T} \\
\end{array}\right).
\end{equation}
The $\tau$-matrix is now purely imaginary.
In other words the magnetic mixing is recast as a kinetic mixing. One can easily check that the effective visible electric charge has remained
the same $-\theta_{12}/(2\pi)$ as before the duality transformation.

\noindent {\bf Example 3:} Finally let us investigate a situation with a purely hidden $\theta_{22}$-term but in addition with
kinetic mixing $\chi\neq 0$ ($\theta_{11}=\theta_{12}=0$).
Again we are interested in the interactions of an electron with a hidden monopole.
Plugging everything into our formula we find
\begin{equation}
E_{\rm Int}=-\frac{1}{4\pi r}\frac{\chi\theta_{22}e_{1}e_{2}}{2\pi(1-\chi^2)},
\end{equation}
and therefore an effective charge $\chi\theta_{22} e_{2}/(2\pi e_{1})$. This can be understood as follows. First the hidden
monopole acquires a hidden electric charge $-\theta_{22}/(2\pi)$.
This then leads to a visible electric charge via kinetic mixing
resulting in an additional multiplicative factor $-\chi e_{2}/e_{1}$.
Moreover, it is straightforward to check that a hidden electric particle interacts with the electron
in exactly the same way as in Example 1, i.e. it has an effective charge  $-\chi e_{2}/e_{1}$.

\section{Generating $\theta$-terms}\label{radiative}

From the viewpoint of a low-energy Lagrangian all terms in Eq.~\eqref{lagrangian} are allowed by the symmetries.
In this section we want to show how they can be generated in the first place from quantum effects due to massive particles of the microscopic high-energy theory.
In particular we will focus on the case when one (or both) of the U(1)s are embedded into a non-abelian gauge group, because this is a setting
where we naturally have magnetic monopoles ('t Hooft-Polyakov monopoles) in addition to electrically charged particles.

\subsection{Kinetic and magnetic mixing in U(1)$\times$SU(2)}
Consider an effective Lagrangian for an U(1)$\times$SU(2) theory,
\begin{eqnarray}
\label{lag}
\nonumber
{\mathcal L}\!\!&=&\!\!
-\frac{1}{4e^{2}_{1}}F^{\mu\nu}F_{\mu\nu}-\frac{1}{4e^2_{2}}G^{a,\mu\nu}G^{a}_{\mu\nu}-\frac{1}{2}D^{\mu}\Phi^{a}D_{\mu}\Phi^{a}-V((\Phi^{a})^2)
-\frac{1}{2M}\Phi^{a}G^{a,\mu\nu}F_{\mu\nu}
\\
&&-\frac{\theta_{11}}{32\pi^2}F^{\mu\nu}\tilde{F}_{\mu\nu}
-\frac{\theta_{22}}{32\pi^2}G^{a,\mu\nu}\tilde{G}^{a}_{\mu\nu}-\frac{1}{16\pi^2M _{\theta}}\Phi^{a}G^{a,\mu\nu}\tilde{F}_{\mu\nu}.
\end{eqnarray}
Here $F^{\mu\nu}$ and $G^{a,\mu\nu}$ are the field strengths of U(1) and SU(2).
The adjoint Higgs which breaks SU(2) to U(1) is denoted by $\Phi^{a}$.
In this effective Lagrangian the interactions between the two gauge sectors are described by the two dimension five operators (last term in the first and second line).
After spontaneous symmetry breaking they result in the kinetic and magnetic mixing terms, respectively. Importantly we will show how these higher dimensional
operators are generated from integrating out a massive field coupled to both sectors.

For a suitable form of the potential $V$, the field $\Phi^{a}$ acquires
a vacuum expectation value (vev). For simplicity we will take it to lie in the $3$-direction.
\begin{equation}
\label{higgsvev}
\langle \Phi^{a}\rangle=(0,0,v).
\end{equation}
Then the $1$- and $2$-components of the gauge fields become massive and the $3$-component provides for the remaining U(1) field, $W^{3,\mu}=A^{\mu}_{2}$, with
gauge field strength, $G^{3,\mu\nu}=F^{\mu\nu}_{2}$.
Using this identification, inserting the vev and retaining only the massless fields, Eq.~\eqref{lag} then becomes equal to Eq.~\eqref{lagrangian} with
\begin{equation}
\chi=\frac{v}{M},\quad \theta_{12}=\frac{v}{M_{\theta}}.
\end{equation}

Let us now see how both the kinetic mixing $\chi$ as well as the magnetic mixing $\theta_{12}$ can indeed be generated naturally by including the quantum effects of
a heavy particle $\Psi$ coupled to both gauge groups (for convenience the U(1) charge is taken to be $1/2$),
\begin{equation}
{\mathcal L}_{\Psi}\!\!=\!\!{\rm i}\bar{\Psi}_{i}\gamma^{\mu}(\partial_{\mu}\mathbf{1}_{ij}+\ii A_{1, \mu}\frac{\mathbf{1}_{ij}}{2}+i W^{a}_{\mu}t^{a}_{ij})\Psi_{j}
-m_{1}\bar{\Psi}\Psi-\ii m_{2}\bar{\Psi}\gamma^{5}\Psi-h_{1}\Phi^{a}\bar{\Psi}_{i}t^{a}_{ij}\Psi_{j}-\ii h_{2}\Phi^{a}\bar{\Psi}_{i}t^{a}_{ij}\gamma^{5}\Psi_{j}.
\end{equation}
Here the $t^{a}=\sigma^{a}/2$ are the generators of SU(2), the U(1) and SU(2) gauge fields are $A_{1, \mu}$ and $W^{a}_{\mu}$,
$h$ is a Yukawa coupling between the Higgs field $\Phi^{a}$ and $\Psi$ and $m$ is a mass term. In order to include CP violating
effects we have allowed both the mass term and the Yukawa coupling to be complex with the imaginary part coupled by a $\gamma^{5}$.

\begin{figure}
\begin{center}
\scalebox{0.7}[0.7]{\fcolorbox{white}{white}{
  \begin{picture}(258,92) (111,-82)
    \SetWidth{1.0}
    \SetColor{Black}
    \Arc[arrow,arrowpos=0.375,arrowlength=5,arrowwidth=2,arrowinset=0.2](240,-29)(45.255,135,495)
    \Arc[arrow,arrowpos=0.88,arrowlength=5,arrowwidth=2,arrowinset=0.2](240,-29)(45.255,135,495)
    \Photon(112,-29)(192,-29){7.5}{4}
    \Photon(288,-29)(368,-29){7.5}{4}
    \Vertex(195,-30){6}
    \Vertex(195,-30){6}
    \Vertex(285,-30){6}
    \Text(125,-12)[c]{\scalebox{1.7}[1.7]{$A^{\mu}_{1}$}}
    \Text(362,-12)[c]{\scalebox{1.7}[1.7]{$A^{\mu}_{2}$}}
  \end{picture}
  }}
\end{center}
\caption{Feynman diagram for a contribution of a particle charged under both gauge groups to the kinetic mixing.} \label{loopfig}
\end{figure}
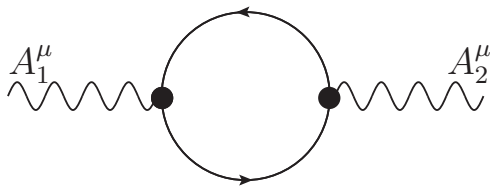

In the above Higgs background, $\Psi$ acquires a mass matrix
\begin{equation}
\label{massmatrix}
\left(
   \begin{array}{cc}
     m_{1}\mathbf{1}+\ii m_{2}\gamma^{5} & 0 \\
     0 & m_{1}\mathbf{1}+\ii m_{2}\gamma^{5} \\
   \end{array}
 \right)
+\frac{1}{2}\left(
    \begin{array}{cc}
      h_{1}v\mathbf{1}+\ii h_{2}v\gamma^{5} & 0 \\
      0 & -(h_{1}v\mathbf{1}+\ii h_{2}v\gamma^{5}) \\
    \end{array}
  \right).
\end{equation}
First we note that due to the spontaneous breaking of the SU(2), the two components of
$\Psi$ have different masses. This invalidates the naive argument
that the diagram shown in Fig.~\ref{loopfig} is $\sim {\rm
Tr}(t^{a})=0$, because the $\Psi$ propagators are no longer
proportional to the unit matrix in SU(2) space. Therefore, we can
indeed have non-vanishing kinetic mixing if the SU(2) is broken
spontaneously. The same reason allows for a non-vanishing $\theta_{12}$.

We now compute the kinetic mixing.
For simplicity let us start with a situation where all masses are real, i.e. $m_{2}=h_{2}=0$
and $|m|=|m_{1}+\ii m_{2}|=|m_{1}|$, $|h|=|h_{1}+\ii h_{2}|=|h_{1}|$.
Under the unbroken $W^{3}_{\mu}\equiv A_{2,\mu}$ field
the first component, $\Psi_{1}$ has charge $1/2$ and mass $m_{1}+h_{1}v/2$ whereas the second
component, $\Psi_{2}$, has charge~$-1/2$ and mass $m_{1}-h_{1}v/2$.
A standard calculation of the loop diagram shown in Fig.~\ref{loopfig} then gives,
\begin{equation}
\label{loopexp}
\chi=\frac{e_{1} e_{2}}{24\pi^2}\log\left(\frac{m_{1}+h_{1}v/2}{m_{1}-h_{1}v/2}\right)\approx \frac{e_{1} e_{2}}{24\pi^2}\frac{hv}{m_{1}}.
\end{equation}
In the last step we have expanded for $h_{1}v\ll m_{1}$.
Restoring the correct index structure for gauge
invariance, we recover exactly the last term in
the first line of Eq.~\eqref{lag} with
\begin{equation}
\frac{1}{M}=\frac{e_{1}e_{2}}{24\pi^2}\frac{h_{1}}{m_{1}}.
\end{equation}

In the next step we want to take the masses and Yukawa couplings to be complex as in the full expression Eq.~\eqref{massmatrix}.
In order to reduce everything to the previous calculation we remove the complex phase by performing chiral rotations.
The complex phase of $m$ can be removed by a transformation,
\begin{eqnarray}
\label{trans1}
&&\Psi_{i}\rightarrow\Psi^{\prime}_{i}=\exp(-\ii \alpha \gamma^{5})\Psi_{i},
\\\nonumber
&&\,\alpha=\frac{1}{2}{\rm Arg} (m_{0}+im_{1})=\frac{1}{2}{\rm Arg}(m).
\end{eqnarray}
These chiral transformations are anomalous, and thus while the classical Lagrangian is invariant, the effect of the anomaly forces the effective Lagrangian to
pick up an additional $\theta$-term,
\begin{eqnarray}
&&\theta_{11}\rightarrow \theta^{\prime}_{11}=\theta_{11}+{\rm Arg}(m),\quad\theta_{22}\rightarrow\theta^{\prime}_{22}=\theta_{22}+{\rm Arg}(m),
\\\nonumber
\label{mass1}
&&m=m_{1}+i m_{2}\rightarrow m^{\prime}=m^{\prime}_{1}=|m|, \\\nonumber
&&h=h_{1}+\ii h_{2}\rightarrow h^{\prime}=h\exp\left(-{\rm Arg}(m)\right).
\end{eqnarray}
$\theta_{12}$ remains unaffected by the transformation because the transformation acts equally on $\Psi_{1}$ and $\Psi_{2}$ which have, however
opposite charge under $A^{\mu}_{2}$ so their contributions cancel. Stated in the original SU(2) language the transformation \eqref{trans1} respects the SU(2)
symmetry and the mixed term in the anomaly vanishes because it contains the trace over a generator $t^{a}$ of SU(2).

After spontaneous symmetry breaking by a vev $v$ of the adjoint Higgs field the complex Yukawa couplings generate a new phase. This phase
can again be removed by chiral rotations. However, this time we have to rotate $\Psi_{1}$ and $\Psi_{2}$ in opposite directions,
\begin{eqnarray}
\label{trans2}
&&\Psi^{\prime}_{1}\rightarrow\Psi^{\prime\prime}_{1}=\exp(-\ii \alpha^{\prime} \gamma^{5})\Psi^{\prime}_{1},
\quad \Psi^{\prime}_{2}\rightarrow\Psi^{\prime\prime}_{2}=\exp(\ii \alpha^{\prime} \gamma^{5})\Psi^{\prime}_{2},
\\\nonumber
&&\,\alpha^{\prime}=\frac{1}{2}{\rm Arg}\left(|m|+\frac{h^{\prime}v}{2}\right)=\frac{1}{2}{\rm Arg} \left(|m|+\frac{h^{\prime}_{1}v+\ii h^{\prime}_{2}v}{2}\right)
=-\frac{1}{2}{\rm Arg} \left(|m|-\frac{h^{\prime}_{1}v+\ii h^{\prime}_{2}v}{2}\right).
\end{eqnarray}
Because the rotations go in opposite directions $\theta_{11}$ and $\theta_{22}$ remain unaffected but
\begin{eqnarray}
\nonumber
&&\theta_{12}\rightarrow \theta^{\prime}_{12}
=\theta_{12}+{\rm Arg}\left(|m|+\frac{h^{\prime}v}{2}\right)
=\theta_{12}+\frac{1}{2}{\rm Arg}\left(m+\frac{hv}{2}\right)-\frac{1}{2}{\rm Arg}\left(m-\frac{hv}{2}\right),
\\\nonumber
&&\qquad\qquad\qquad\qquad\qquad\qquad\qquad\quad\,\,\,\,\,\approx \theta_{12}+\frac{h_{2}v}{2|m|},
\\
\label{mass2}
&&|m|\pm\frac{h^{\prime}v}{2}\rightarrow \left|m\pm \frac{hv}{2}\right|.
\end{eqnarray}
In the second line we have assumed $m$ to be real and $|hv|\ll|m|$. The proportionality to $h_{2}v$ emphasizes again that $\theta_{12}$ arises only if
the SU(2) symmetry is spontaneously broken. In the language of Eq.~\eqref{lag} we have
\begin{equation}
\frac{1}{M_{\theta}}=\frac{h_{2}}{2|m|}.
\end{equation}

After these rotations all masses are real and we can perform the calculation of the kinetic mixing as before but with the masses given by Eq.~\eqref{mass2}.

Overall, our calculation shows that magnetic mixing arises as a natural partner of kinetic mixing for complex fermion masses and Yukawa couplings.
We note that $\theta_{11}$ and $\theta_{22}$, being just the phase of the fermion mass term, could be naturally of order $1$.
On the other hand the magnetic mixing term $\theta_{12}$ could be naturally small if $|m|\gg |hv|$ with both $\theta_{12}$
and the kinetic mixing of the same order, $\theta_{12}\sim\chi\sim |hv|/|m|$.

\subsection{Electromagnetic mixing from underlying SU($N$) dynamics}

In the previous section we have calculated the kinetic and magnetic mixing terms for a scenario with a priori independent U(1) and SU(2)
gauge groups.
Here, we consider a scenario where the entire low-energy U$(1)^{k}$ structure of the visible
and the hidden sectors descend from a non-abelian simple gauge group.

Although both scenarios are in principle complementary, it is instructive to note that the setup of the previous subsection can be straightforwardly
embedded into a single gauge group. One possibility would be to take an SU(3) gauge group with a fundamental fermion and break it by
an adjoint vev $\sim (-a, a/2+v,a/2-v)$ with $a\gg v$. At the scale $a$ the symmetry is broken to U(1)$\times$SU(2) and the fermion splits
into a singlet charged under U(1) and a charged SU(2) doublet. The singlet does not contribute to the kinetic mixing whereas the doublet is exactly
the matter representation of the previous subsection. At the scale $v$ the SU(2) subgroup is then broken to U(1) as in the previous section.

For concreteness we assume a microscopic description
in terms of an SU($N$) gauge theory with a generic matter content and an adjoint Higgs field which breaks
SU($N$)$\to$U(1)$^{k}$, where $k=N-1$. Upon integrating out massive degrees of freedom,
the coupling constant of the `unified' SU($N$) theory splits into the $k \times k$ matrix
of U(1) couplings $e_{ab}$ below the unification scale set by vevs of the adjoint. At one-loop we can write a general
expression
\begin{eqnarray}
\left(\frac{4 \pi}{e^2}\right)_{ab} \, = \, \frac{7}{4\pi} \sum_{V} \Tr \left(t_a^Vt_b^V \log \frac{M_V}{\Lambda} \right) &-&
\frac{2}{3\pi} \sum_{F} \Tr \left(t_a^F t_b^F \log \frac{M_F}{\Lambda} \right) \nonumber\\
&-& \frac{1}{12\pi} \sum_{S} \Tr \left(t_a^S t_b^S \log \frac{M_S}{\Lambda} \right)
\label{threshold}
\end{eqnarray}
where $\Lambda$ is the dimensional transmutation scale of the underlying SU($N$) theory, and the sums are over all massive vector,
Dirac fermion and real scalar degrees of freedom which were integrated out.
The matrices $t_a$ are the appropriate generators of the unbroken U(1) factors. For example, if the microscopic theory is
U($N$) (if desired the overall U(1) factor can be decoupled later on) broken down to U(1)$^N$, the U(1) generators in the
adjoint and in the fundamental representations are given by (repeated indices not summed over)
\begin{equation}
(t_a^{\rm adj})^{bd}_{ce} = \frac{1}{\sqrt{2}} \delta_e^b\delta_c^d(\delta_a^b -\delta_c^a) \ , \qquad
(t_a^{\rm fund})^{b}_{c} = \frac{1}{\sqrt{2}} \delta_c^b\delta_a^b \ , \qquad a=1, \ldots, N.
\end{equation}
The matrix of low-energy U(1) couplings in \eqref{threshold} follows from the
general expression for one-loop threshold corrections derived by Weinberg in \cite{Weinberg:1980wa}
and we are using the result converted to the $\overline{\rm DR}$ scheme \cite{Martin:1993yx}.

Contributions to $\theta$ arise only from the complex masses of fermions (as in the previous subsection),
\begin{equation}
\label{threshold2}
\left(\frac{\theta}{2\pi}\right)_{ab} = - \frac{1}{\pi} \sum_{F} \Tr \left(t_a^F t_b^F {\rm Arg} \left( \frac{M_F}{\Lambda}\right) \right),
\end{equation}
where the sum is only over Dirac matter fermions. Neither scalars nor vector fields play a role as they do not contribute to the axial anomaly.

We now make two observations. First, it is obvious that \eqref{threshold} describes a generic matrix of coupling constants
with non-vanishing non-diagonal entries. These give the kinetic mixing effects induced in the low-energy theory from integrating
out heavy states of the high-energy theory. Second, we note that \eqref{threshold} also knows about the magnetic mixing given in Eq.~\eqref{threshold2}.
In the original derivation of \eqref{threshold} it was assumed that all the masses are real and positive. Otherwise they should appear in
Eq.~\eqref{threshold} as $|M|$. However, in the presence of supersymmetry and in particular if the underlying theory is ${\cal N}=2$ supersymmetric
the entire $\tau$-matrix of complexified couplings has an analytic structure. It is given by the second derivative of the Seiberg-Witten prepotential
which is a holomorphic function of its variables. In this case $\tau$ can be directly read off from the right hand side of Eq.~\eqref{threshold} with in general
complex $M$s.
For ${\cal N}=1$ supersymmetry we can easily see that the complex phase contributions from the second and third terms in
Eq.~\eqref{threshold}, i.e. the contributions from chiral matter, combine to give Eq.~\eqref{threshold2}.

To summarize we conclude that magnetic mixing of the low-energy abelian theory is generated alongside with the kinetic mixing.

As an example, let us show how this works explicitly for a Seiberg-Witten ${\cal N}=2$ QCD with $N$ colours and $N_f$ fundamental
hypermultiplets. In the context of the U($N$) gauge theory spontaneously broken to U(1)$^N$ in the Coulomb phase,
the $N \times N$ matrix of effective complexified couplings
of the low-energy theory reads
\begin{equation}
\hat{\tau}_{ab} \,=\, \delta_{ab}\, \frac{i}{2\pi}\, \left(\sum_{c \neq a}^N \log \left(\frac{v_a-v_c}{\Lambda}\right)^2 -
N_f \log \left(\sqrt{2} \frac{v_a}{\Lambda}\right)\right) \, -\,
(1-\delta_{ab})\, \frac{i}{2\pi}\, \log \left(\frac{v_a-v_c}{\Lambda}\right)^2
\label{SWtauab}
\end{equation}
In this formula $v_a$ are the vevs of the adjoint Higgs, $\langle \Phi \rangle = {\rm diag}(v_1, \ldots, v_N)$,
hence the masses of vector bosons (and their superpartners) are $|v_a-v_b|$
while the fundamental flavour masses for each U(1) factor are $|\sqrt{2} v_a|$.
In order to describe SU($N$)$\to$U(1)$^{N-1}$ we need to
decouple the overall U(1) factor from the U($N$) theory. The resulting $(N-1) \times (N-1)$ $\tau$-matrix in the
low-energy abelian theory is obtained from \eqref{SWtauab} via
\begin{equation}
\tau_{IJ} \,=\, \hat{\tau}_{IJ} - \hat{\tau}_{IN} -\hat{\tau}_{NJ} + \hat{\tau}_{NN} \ , \qquad I,J=1, \ldots, N-1
\label{SWtauij}
\end{equation}

Equations \eqref{SWtauab}, \eqref{SWtauij} describe the perturbative interaction pattern between the gauge factors of the
U(1)$^{N-1}$ low-energy effective theory which arises from the Seiberg-Witten SU($N$) SQCD. They are presented in the
$\overline{\rm DR}$ regularisation scheme with a careful account of the threshold corrections. In this form they were first
obtained in \cite{Dorey:1996bn} for $N=2$ colours and later in \cite{Argyres:1999ty} (whose notation we followed here)
in the context of the general SU($N$) Seiberg-Witten theory.

It follows from the Seiberg-Witten construction \cite{Seiberg:1994rs,Seiberg:1994aj} that
these one-loop expressions are exact to all orders in perturbation theory
and receive only non-perturbative (multi-instanton) corrections which were computed directly in~\cite{Dorey:1996hu,Khoze:1998gy}.
In the case of more general non-supersymmetric settings, the above results represent the leading order
perturbative expression for the effective $\tau$ matrix. To summarise results of this subsection, we conclude that in rather general settings
where several U(1) factors of the abelian low-energy theory arise from the same non-abelian gauge group, kinetic and magnetic
mixing effects arise naturally (the latter require CP-violation) and are governed by a simple one-loop expression  for
$\tau_{IJ}$.

\section{Conclusions}\label{conclusions}
We have argued that when a low-energy theory contains two or more U(1) factors
there is a natural partner of kinetic mixing which we call magnetic mixing.
Kinetic mixing manifests itself as a non-diagonal kinetic term for the gauge fields, $\chi F_{1}F_{2}$,
magnetic mixing is the corresponding mixing in the $\theta$-terms, $\theta_{12} F_{1}\tilde{F}_{2}$.
Both terms are naturally generated upon integrating out heavy degrees of freedom -- in the latter case
in presence of CP violating phases -- of an underlying theory. They occur independently of whether
the U(1) factors are fundamental or arise from a single, simple non-abelian gauge group.

Consider an application where one of the U(1)s corresponds to ordinary electromagnetism and the other belongs to a hidden sector.
Then both mixing effects generate interactions between the visible sector and hidden matter. More precisely, they induce
non-integer (generically small) electric charges for hidden sector states. In the case of kinetic mixing, visible mini-electric charges
are induced on charged hidden matter. Similarly, magnetic mixing induces visible mini-electric charges on the magnetic monopoles of the hidden sector.
When the carriers of these mini-charges are light they can be probed by current and future low-energy
experiments~\cite{Jaeckel:2009dh,Gies:2006hv,Badertscher:2006fm,Gninenko:2006fi,Ahlers:2007rd,Cameron:1993mr,Gies:2009wx,Gies:2006ca,Zavattini:2007ee,Dobrich:2009kd,Melchiorri:2007sq,Ahlers:2009kh,Gluck:2007ia}.

\section*{Acknowledgements}
The authors would like to thank Steve Abel and Frank Krauss for discussions.

\begin{appendix}
\section{${\rm Sp}(2k,\mathbf{Z})$ duality in abelian gauge theory}
As pointed out in the main text, in a U$(1)^k$ theory the duality group Sp$(2k,\mathbf{Z})$ acts on the coupling matrix $\tau$ and on the electric and magnetic quantum numbers. This ``Siegel modular group'' consists of $2k\times 2k$ matrices $M$ with integer entries which leave a symplectic inner product invariant:
\begin{equation}
M^T\left(\begin{array}{cc} 0 & \mathbf{1}_k \\ -\mathbf{1}_k & 0
\end{array}\right) M=\left(\begin{array}{cc} 0 & \mathbf{1}_k \\
-\mathbf{1}_k & 0 \end{array}\right).
\end{equation}
$M\in{\rm Sp}(2k,\mathbf{Z})$ may be written in terms of $k\times k$ block matrices as
\begin{equation}
M=\left(\begin{array}{cc} A & B \\ C & D \end{array}\right),
\end{equation}
with $A,B,C,D$ satisfying
\begin{equation}
\begin{split}
AB^T-BA^T=CD^T-DC^T&=A^TC-C^TA=B^TD-D^TB=0,
\\ A^TD-C^TB&=AD^T-BC^T=\mathbf{1}_k.
\end{split}
\end{equation}
$M$ acts on the coupling matrix as
\begin{equation}
M:\,\tau\,\rightarrow\,(A\tau+B)(C\tau+D)^{-1}
\end{equation}
and on electric and magnetic quantum numbers according to
\begin{equation}\label{qnaction}
M:\,\left(\begin{array}{c} n^{e} \\ n^{m} \end{array}\right)
\,\rightarrow\, M \left( \begin{array}{c} n^{e} \\ n^{m} \end{array}\right).
\end{equation}

Sp$(2k,\mathbf{Z})$ is generated by discrete $S$ and $T$ transformations. $S$ is the electric-magnetic duality implemented via an inversion of the coupling matrix,
\begin{equation}
S:\,\tau\,\rightarrow\,-\tau^{-1}.
\end{equation}
$T$ transformations are integer shifts of the ${\rm Re}\,\tau_{IJ}$ (corresponding to  changing the $\theta_{IJ}$ by $2\pi$):
\begin{equation}
T^{KL}:\,\tau_{IJ}\,\rightarrow\,\tau_{IJ}+\delta^K_I\delta^L_J+\delta^L_I\delta^K_J-\delta_{IJ}\delta^K_I\delta_J^L\qquad(1\leq K\leq L\leq k\text{, no summation}).
\end{equation}
$S$ is explicitly given by
\begin{equation}
S=\left(\begin{array}{cc} 0 & -\mathbf{1}_k \\ \mathbf{1}_k & 0
\end{array}\right).
\end{equation}
From this and Eq.~\eqref{qnaction} it is obvious that $S$ exchanges electric and magnetic quantum numbers.

\end{appendix}

\end{document}